\documentstyle[prl,aps]{revtex}
\input epsf

\tighten
\begin{document}

\newcommand{\be}{\begin{equation}} 
\newcommand{\ee}{\end{equation}} 
\newcommand{\ba}{\begin{array}} 
\newcommand{\ea}{\end{array}}

\title{Neutrino Zero Modes on Electroweak Strings}

\author{
Glenn Starkman, Dejan Stojkovic and Tanmay Vachaspati}

\address
{Department of Physics,
Case Western Reserve University,
10900 Euclid Avenue, Cleveland, OH 44106-7079, USA}

\wideabs{
\maketitle

\begin{abstract}
\widetext

Zero  modes of massive standard model fermions have been found
on electroweak Z-strings.
A zero mode solution for a massless left-handed neutrino is also known,
but was thought to be non-normalizable.
Here we show that although this mode is not discretely normalizable, 
it {\em is} delta-function normalizable and
the correct interpretation of this solution is 
within the framework of the continuum spectrum. 
We also analyze an extension of the standard model including right-handed neutrinos
in which neutrinos have Dirac masses,
arising from a Yukawa coupling to the usual SU(2) Higgs doublet,
and  right-handed Majorana masses.
The Majorana mass terms are taken to be spatially homogeneous
and are presumed to arise from the vacuum expectation value of some field 
acquired in a phase transition well above the electroweak phase transition.
The resulting zero energy equations have a discrete zero mode.

\end{abstract}
}

\narrowtext

\section{Introduction}

The interactions of fermions with solitons has led to the discovery
of many novel and important field theoretic phenomena such as
fractional charges \cite{JacReb} and quantum anomalies \cite{Adler1969,BellJackiw1969,tHo76}. 
Fermionic zero modes on strings \cite{Nohl,deVega,Jackiw} may play 
an important role in cosmology since they can be responsible for 
string superconductivity \cite{Witten} and vortons \cite{vortons}.
Although the standard electroweak theory does not contain any topological
defects, ``embedded defects" such as Z-strings are 
possible \cite{Nambu,Barriola,Zstring}. 
Fermionic (quark and leptonic) zero modes on electroweak Z-strings
have been studied in the literature \cite{Perkins,Vachaspati,Oakninetal}
and can provide linked configurations of strings
with anomalous values of baryonic charge and other quantum numbers
\cite{Vachaspati,VacFie94}. Quark and leptonic zero modes also
have an important effect on the stability of the 
Z-string \cite{Naculich,Hong,Groves}. 

The work on fermionic zero modes on Z-strings has so far 
considered massless neutrinos. The conclusion 
of these studies has been \cite{Nohl,deVega,Jackiw,Vachaspati,Oakninetal} that there are no
normalizable zero energy solutions for massless particles in the 
vortex background. Some attention had been given to massive neutrinos before, 
but in the framework of strings in Grand Unified models \cite{Davis1,Davis2}.
A reconsideration of both massive and massless neutrino zero modes on 
electroweak strings is needed. This is the aim of the present paper.

In Section 2, we present the electroweak standard model Lagrangian for 
leptons and find the neutrino equations of motion in a Z-string background.
We find the solutions of these equations of motion,
and show that a zero mode does exist as part of the continuum.
In Section 3 we present the standard extension of the electroweak
Lagrangian with a massive neutrino and derive the neutrino equations of motion.
We solve the equations of motion in two asymptotic regimes 
(small and large $r$) and discuss the number of well-behaved solutions. 
We finally solve the equations numerically to obtain the
zero mode solution. 

There are several index theorems in the literature which specify the
number of zero modes in a given Lagrangian. The index theorem in 
\cite{Weinberg} does not apply to a massive neutrino case because it
assumes that the source of all fermionic masses is the interaction with a 
single Higgs field (which is not true for the right-handed neutrino Majorana
mass). The generalized index theorem in \cite{Lazarides1} and \cite{Lazarides2} 
also does not apply because it uses an ansatz for the spinor components
not general enough to cover our case. The generalized index theorem in
\cite{Davis1} applies to the massive neutrino with a Dirac mass and either
a left- or a right-handed Majorana mass but not both.



\section{The Standard Model Neutrino}

We consider first the Glashow-Weinberg-Salam electroweak standard model, 
and focus on the interactions of a  single lepton family. 
The fermion content is a left-handed SU(2) lepton doublet, 
$\Psi\equiv(\nu_L, e^-_L)^T$ 
and a right-handed electron singlet $e^-_R$.
Electrons acquire their (Dirac) mass from Yukawa coupling to the SU(2) doublet Higgs field $\Phi$, 
while neutrinos are massless. (There is no right-handed neutrino to participate in
a Dirac mass, and there is assumed to be no left-handed Majorana mass, since
there is no fundamental Higgs triplet in the model.) 
The Lagrangian of the theory is:
\begin{eqnarray}
\label{LSM}
{\cal L}_{SM}=& \hspace{-.5cm} -{1\over4}W^a_{\mu\nu}W^{a\mu\nu}-{1\over4}
F_{\mu\nu}F^{\mu\nu}+
     \left(D_\mu\Phi\right)^\dagger\left(D^\mu\Phi\right) - \nonumber\\
-&\lambda\left(\Phi^\dagger\Phi-\eta^2\right)^2  
-i{\bar\Psi}\gamma^\mu D_\mu\Psi - i\overline{e_R}\gamma^\mu D_\mu e_R +  \\
+& \hspace{-3cm} h'\left(\overline{e_R}\Phi^\dagger\Psi+ 
          {\bar\Psi}\Phi e_R\right) \nonumber\ . 
\end{eqnarray} 
\noindent Here 
\begin{eqnarray}
D_\mu \Psi & \equiv & \left(\partial_\mu-i{g\over2}\tau^a W^a_\mu +
                i{{g'}\over2}B_\mu\right)\Psi \ ,\nonumber \\
D_\mu e_R  & \equiv &  \left(\partial_\mu+
                ig'B_\mu\right)e_R  \\
D_\mu \Phi & \equiv & \left(\partial_\mu-i{g\over2}\tau^a W^a_\mu -
                i{{g'}\over2}B_\mu\right)\Phi \ , \nonumber
\end{eqnarray}
with
\begin{eqnarray}
W^a_{\mu\nu} & \equiv & \partial_\mu W^a_\nu-\partial_\nu W^a_\mu+
                           g\epsilon^{abc} W^b_\mu W^c_\nu \ , \\
F_{\mu\nu}   & \equiv & \partial_\mu B_\nu- \partial_\nu B_\mu  \nonumber
\end{eqnarray}
where $\tau^a$ are the Pauli matrices.
As is conventional, we define
\begin{eqnarray}
Z_\mu \equiv & \cos\theta_w W^3_\mu-\sin\theta_w B_\mu   \nonumber \\
A_\mu \equiv & \sin\theta_w W^3_\mu+\cos\theta_w B_\mu    \nonumber
\end{eqnarray}
where $\theta_w$ is the weak mixing angle.
The fermionic equations of motion for Lagrangian (\ref{LSM}) are
\begin{eqnarray}
i\gamma^\mu D_\mu \Psi & = & h'\Phi e_R \nonumber \\
i\gamma^\mu D_\mu e_R & = & h'\Phi^\dagger \Psi_L  \ .
\end{eqnarray}

In the Z-string ansatz, $A^\mu = W^1_\mu =W^2_\mu =0$,
$\Phi=\left( 0, \phi \right)^T$. 
In cylindrical coordinates $(r,\theta,z)$ the gauge and Higgs fields
take the form
\be
qZ^\mu=\left(0,-{{v(r)}\over r}{\vec e}_\theta\right),
\quad \phi=\eta f(r)e^{i\theta}
\label{zstringfields}
\ee
where $q \equiv \alpha /2 = \sqrt{g^2+g'^2} /2$, and 
$v(r)$ and $f(r)$ are the Z-string profile functions.

With this ansatz, the neutrino equation of motion is: 
\be
\label{eqn:SMnu}
 i\gamma^\mu D_\mu \nu_L = 0
\ee
A representation of the  Dirac matrices in cylindrical coordinates is


\be
\label{Gamma}
 \gamma^r=\pmatrix{  
0 & e^{-i\theta} & 0 & 0 \cr 
-e^{i\theta} & 0 & 0 & 0 \cr
0 & 0 & 0 & -e^{-i\theta}\cr
0 & 0 & e^{i\theta} & 0 }, 
\ee
\be
\gamma^\theta=
\pmatrix{  0 & -ie^{-i\theta} & 0 & 0 \cr 
-ie^{i\theta} & 0 & 0 & 0 \cr
0 & 0 & 0 & ie^{-i\theta} \cr
0 & 0 & ie^{i\theta} & 0 }
\ee

\[ \!\!\!\!\!\! \gamma^0=\pmatrix{\tau^3&0\cr 0&-\tau^3\cr},\ \ \ \gamma^z=\pmatrix{
0&1\cr -1&0\cr}, \ \ \ \gamma^5=\pmatrix{0&1\cr 1&0} \]
In this representation, a left-handed Dirac fermion has the form: 
$\nu^T_L =(\alpha ,\beta , -\alpha , -\beta )$.

A general solution to equation (\ref{eqn:SMnu}) can be written as a 
superposition of all modes with definite winding number, 
\begin{eqnarray} \label{SMansatz}
\alpha  =&\sum_{m=-\infty}^{\infty} \alpha_m(r) e^{ik_z z -i\omega t + 
                     i m\theta} \\
\beta  =&\sum_{m=-\infty}^{\infty}  i \beta_m(r) e^{ik_z z -i\omega t + 
                     i m\theta} \nonumber
\end{eqnarray}
$\alpha_m$ and $\beta_m$ are, in general, complex functions of $r$.
We are interested in zero energy solutions to equation (\ref{eqn:SMnu}) 
such that all spinor components fall off outside the string core 
(large $r$) fast enough to be normalizable and are well-behaved at the 
origin (small $r$). 

After setting $k_z=0$, we get a set of
two recursive equations for the coefficients of $\alpha$ and $\beta$:
\begin{eqnarray} 
\label{smn}
\omega \alpha_{m-1} + \beta'_m+\frac{m+v}{r} \beta_m = &0  \\
-\omega \beta_{m} + \alpha'_{m-1}-\frac{m-1+v}{r}\alpha_{m-1} = & 0 \nonumber 
\end{eqnarray}
Let us consider the case $m=0$. For large $r$, $v\sim 1$, $f\sim 1$ and the closed system of equations is:
\begin{eqnarray} \label{large r}
\omega \alpha_{-1} + \beta'_0+\frac{1}{r} \beta_0 = & 0  \\
-\omega \beta_{0} + \alpha'_{-1}= & 0 \nonumber
\end{eqnarray}
Eliminating $\alpha_{-1}$ from the system we get Bessel's equation for $\beta$:
\be
\beta''_0+\frac{1}{r} \beta'_0+\beta_0 (\omega^2 - \frac{1}{r^2})=0
\ee
the general solution of which is given by
\be \label{wave}
\beta_0(\omega r) = A \sqrt{\omega} J_1(\omega r) + B \sqrt{\omega} Y_1(\omega r)
\ee
\noindent Here $J_1$ and $Y_1$ are Bessel's functions of the
first and second kind, while $A$ and $B$ are constants.
We see therefore that the $k_z=0$ solutions comprise a continuum labeled by 
$\omega$. Using the normalization relation for Bessel's functions:
\be
\int_0^\infty J_1(\omega r) J_1(\omega' r) r dr = \frac{1}{\omega}
\delta (\omega' - \omega) \ .
\ee
and a similar relation which holds for $Y_1$, 
we see that the solution is properly $\delta$-function normalized.
For $\omega \neq 0$ and $r$ large, (\ref{wave}) is just a cylindrical wave 
$\beta_0 \sim e^{\pm i\omega r}/\sqrt{r}$.
To see what happens with the zero mode, i.e. $\omega =0$, solution, we take
$r$ fixed but large enough for eq. (\ref{large r}) to be valid,
say $r \gg \eta^{-1}$. Thus, the limit $\omega \rightarrow 0$ implies
that $\omega r \rightarrow 0$. Using the asymptotic expansion for Bessel's
function of small argument we get the neutrino wave function:
\be
\beta_0 = \frac{D}{ \sqrt{\omega} r}
\ee
where $D$ is a constant independent of $\omega$ and $r$.

The neutrino equations for small $r$ and $m=0$ are:
\begin{eqnarray} \label{small r}
 \omega \alpha_{-1} + \beta'_0 + v_0 r \beta_0 = & 0  \\
 -\omega \beta_0 + \alpha'_{-1} + \frac{1}{r} \alpha_{-1} =& 0 \nonumber
\end{eqnarray}
where we used $v\sim v_0 r^2$ and $f \sim f_0 r$ near the origin.
Eliminating $\alpha_{-1}$ from the system we get equation for $\beta$:
\be
\beta''_0+\frac{1}{r} \beta'_0+\beta_0 (\omega^2 +2 v_0)=0
\ee
whose solution is given by
\be
\beta_0(\omega r) = C I_0(\sqrt{-\omega^2-2v_0} ~ r) 
\ee

\noindent where $C$ is a constant and $I_0$ is a modified Bessel 
function of the first kind.
Thus, $\beta_0 \sim \sqrt{\omega} C$ near the origin, where $\sqrt{\omega}$ is
introduced in order to have a consistent normalization. This is in agreement
with the result in \cite{Vachaspati}. 

$\beta_0$ is the only nontrivial solution of the system (\ref{smn}) in the
zero mode limit, i.e. $\omega r \rightarrow 0$.
The solutions for all other $\alpha_m$ and $\beta_m$ are 
not valid in the zero mode limit because of 
singular behavior at either small or large $r$ which makes solutions not
even Dirac delta function normalizable. 

If we look at the solution for $\beta_0$ as an isolated zero mode, it is 
obviously not normalizable because the normalization integral diverges 
logarithmically \cite{Vachaspati}. But, we have seen that this state is 
actually part of a continuum spectrum of the theory and is Dirac delta 
function normalizable. Therefore it is a valid zero mode solution.

\section{Beyond the Standard Model}

Recent evidence from the Super Kamiokande experiment \cite{Kamiokande} 
strongly suggests that neutrinos are very light, but not massless.
A simple extension to the standard model which incorporates light neutrino 
masses is to add to each standard model family an $SU(2)_L \times U(1)_Y$ 
singlet right-handed neutrino,
and to incorporate both a Dirac mass  for the neutrino and a Majorana mass
for the right-handed neutrino. The origin of $M_R$ is a vacuum expectation 
value of some $SU(2)$ singlet Higgs field, acquired at some energy scale much 
greater than that of the electroweak scale. $M_R$ can therefore be taken to 
be spatially homogeneous. The modified Lagrangian is

\begin{eqnarray}
\label{Lbeyond}
{\cal L}={\cal L}_{SM} &+& 
       h \left[\bar\Psi i \tau_2 \Phi^\ast \nu_R 
               - \overline{\nu_R}\Phi^T i \tau_2^{\dagger} \Psi\right] \nonumber \\
               &+& {1\over2} \overline{\nu_R^c} M^\ast_R \nu_R +
                               {1\over2} \overline{\nu_R} M_R\nu_R^c  
\end{eqnarray} 
where $\nu^c \equiv  C \bar\nu^T$ and 
$\nu^T_R=(\gamma ,\delta , \gamma ,\delta)$. 
The charge conjugation matrix $C$ in the representation we are working in is: 
\be
C \equiv i \gamma^2 \gamma^0 = \pmatrix{ 0&-1&0&0 \cr 1&0&0&0 \cr
0&0&0&-1 \cr 0&0&1&0} \ .
\ee
The resulting neutrino equations of motion are:
\begin{eqnarray}
\label{neq}
i\gamma^\mu D_\mu \nu_L =& \hspace{-1.6cm} h\phi^\ast \nu_R \\
i\gamma^\mu \partial_\mu \nu_R =& h\phi \nu_L + M_R (\nu_R)^c 
\nonumber 
\end{eqnarray}

The presence of the Majorana mass term, $M_R$, 
prevents us from removing the angular 
dependence from the Dirac equations for the neutrino fields using the standard 
procedure \cite{Perkins,Jackiw}. Instead, we must consider a 
superposition of all modes with definite winding number.
Extending the ansatz (\ref{SMansatz}), we write
\begin{eqnarray} \label{fullansatz}
\alpha  =& \sum_{m=-\infty}^{\infty}  \alpha_m(r) e^{ik_z z -i\omega t + i m\theta} \nonumber \\
\beta  =& \sum_{m=-\infty}^{\infty} i \beta_m(r) e^{ik_z z -i\omega t + i m\theta} \\
\gamma  =& \sum_{m=-\infty}^{\infty}   \gamma_m(r) e^{ik_z z -i\omega t + i m\theta} \nonumber\\
\delta  =& \sum_{m=-\infty}^{\infty}  i \delta_m(r) e^{ik_z z -i\omega t + i m\theta} \nonumber
\end{eqnarray}

After setting $\omega=k_z=0$, we get a set of four 
recursive equations for the coefficients
$\alpha, \beta, \gamma$ and $\delta$
\begin{eqnarray}
\label{rec}
\beta'_m + {(m+v)\over r}\beta_m  = &-h\eta f \gamma_{m} \nonumber  \\
\gamma'_m - {m\over r} \gamma_m - iM_R \gamma^\ast_{-1-m} =& -h\eta f \beta_{m} \\
\alpha'_m - {(m+v)\over r}\alpha_m  =& -h\eta f \delta_{m+2} \nonumber \\
\delta'_m + {m\over r} \delta_m - iM_R \delta^\ast_{1-m} =& -h\eta f \alpha_{m-2} 
\nonumber
\end{eqnarray}

\noindent In the $M_R=0$ limit, the equations are not recursively connected.

System (\ref{rec}) decomposes into two subsystems - ($\beta$, $\gamma$) and ($\alpha$, $\delta$). 
Let us first analyze the ($\beta$, $\gamma$) equations. From (\ref{rec}) we see that
a self consistent set of equations consists of four equations for 
$\beta_m$, $\beta_{-1-m}$, $\gamma_m$ and $\gamma_{-1-m}$.
The case $m=0$ is the most interesting one
because we already know that in the $M_R =0$ case (e.g. electrons), zero 
modes have $m=0$ \cite{Vachaspati}. Setting $m=0$ in the first two equations
of the system (\ref{rec}) we get:

\begin{eqnarray} \label{m=0}
\beta'_0 + {v\over r}\beta_0  = &-h\eta f \gamma_{0}  \nonumber \\
\beta'_{-1} + {(v-1)\over r}\beta_{-1}  = &-h\eta f \gamma_{-1}  \\
\gamma'_0  - iM_R \gamma^\ast_{-1} =& -h\eta f \beta_{0}
\nonumber \\
\gamma'_{-1} + {1\over r} \gamma_{-1} - iM_R \gamma^\ast_{0} =& -h\eta f \beta_{-1}
\nonumber \end{eqnarray}

An analytic solution to the system (\ref{m=0}) could not be found, but we can 
learn something about the structure of solutions by looking at their 
asymptotic behavior.
Neglecting fermion back reaction to the string background,
$v$ and $f$ are just the Nielsen-Olesen profiles and at 
large $r$ behave as $f(r \rightarrow \infty)\rightarrow 1$, 
$v(r \rightarrow \infty) \rightarrow 1$. 
Assuming the asymptotic behavior  $\beta_0 \sim a e^{s r}$, $\gamma_{0} \sim b e^{t r}$, 
$\beta_{-1} \sim c e^{u r}$, $\gamma_{-1} \sim d e^{l r}$
($a$, $b$,$c$ and $d$ are arbitrary complex numbers while $s$, $t$, $u$ and $l$ 
are complex numbers, with a nonpositive real part) we get 
following conditions:

\be
s=t=u^\ast=l^\ast \ \ \ \ \ \mathrm{and} \ \ \ \ \ \
|M_R|^2 s^2 = (|h\eta|^2-s^2)^2
\ee 
The solutions to this fourth order polynomial equation for $s$ are
given by

\begin{equation}
s^2_\pm =
{1\over 2} \left[ |M_R|^2 +  2|h\eta|^2 
                      \pm |M_R| \sqrt{|M_R|^2 + 4|h\eta|^2} \right] 
\end{equation}

All four of these solutions for $s$ are real. Two of these are positive,
giving rise to exponentially growing modes and two are negative, giving
rise to exponentially decaying modes. Only the latter are 
physically acceptable. In the physical limit of $M_R \gg |h\eta|$ we get
$s_+=M_R$ and $s_-=(h\eta)^2/(2 M_R)$, which is the standard ``see-saw"
result --- outside the string core neutrino zero modes decay such that one
decay channel is driven by the heavy neutrino (vacuum) mass eigenstate and 
another one is driven by the light one. In the limit of 
$M_R/(h \eta) \rightarrow 0$ we have the standard result 
$s=\pm |h \eta|$ \cite{Jackiw}.

If we want to match the exponentially decaying large $r$ solutions in a given
pair of equations ($\beta$,$\gamma$) to the solutions at the origin, we must 
know how many solutions for small $r$ are well-behaved for the set of 
$(\beta_0, \gamma_0, \beta_{-1}, \gamma_{-1})$ . By well-behaved we mean 
nonsingular and single-valued. We require not only the functions, but also all 
their derivatives to be single-valued in order to have a well defined solution.
Therefore, for a solution of the form $r^p e^{i q \theta}$ the following must
be true: (i) $p \geq |q|$, (ii) both $p$ and $q$ are even or both are odd.
Neglecting fermion back reaction on the string background, the asymptotic 
behavior of the profile functions is $v\sim v_0 r^2 + v_2 r^4 + \ldots $ 
and $f\sim f_0 r + f_2 r^3 +\ldots $ for small $r$. 
Writing
$\beta_0 \sim a r^s$, $\gamma_{0} \sim b r^t$, 
$\beta_{-1} \sim c r^u$, $\gamma_{-1} \sim d r^l$
($a$, $b$,$c$ and $d$ are arbitrary complex numbers and $s$, $t$, $u$ and $l$ 
are nonnegative real numbers, $s$ and $t$ even, $u$ and $l$ odd) for small $r$ 
we find that the leading orders are
$s=0$, $t=0$, $u=1$ and $l=1$. System (\ref{m=0}) is invariant under the parity
transformation ($r \rightarrow -r$), so all corrections to the solutions 
which are parity violating are zero. We can write the general well behaved 
solution in the form:
\be \label{m=0sol}
\left( \begin{array}{c} \beta_0 \\ \gamma_0 \\ \beta_{-1} \\ 
                     \gamma_{-1} \end{array} \right) 
 = \left( \begin{array}{c}
a_0 r^0 + a_2 r^2 + a_4 r^4 + \ldots \\
b_0 r^0 + b_2 r^2 + b_4 r^4 + \ldots \\
c_0 r^1 + c_2 r^3 + c_4 r^5 + \ldots \\
d_0 r^1 + d_2 r^3 + d_4 r^5 + \ldots 
\end{array} \right) 
\ee

The structure of the system (\ref{m=0}) is such that only three coefficients are
independent (say $a_0$, $b_0$ and $d_0$ are independent, while 
$c_0 = i M_R b_0/2$ and all other coefficient are functions
of the first three) which says that there are three linearly
independent well-behaved solutions. For the system of four linear first order
differential equations we should have four solutions in total, so we conclude
that one solution is not well-behaved, i.e. contains one or more terms such as
$r^p$, $p<0$ and/or $r^p (\log(r))^q$.

With three well-behaved solutions at the origin and two well-behaved solutions 
at infinity there must be at least one solution which is well-behaved 
everywhere. Each of the two well-behaved solutions at infinity matches on to
a unique linear combination of all four solutions at the origin where only one
of them is bad, so there is always one linear combination of the two good
solutions at infinity which does not have any contribution from the single
bad solution at the origin. However, it could happen that there are two
solutions which are well-behaved everywhere if each of the two good solutions
at infinity matches exactly to a linear combination of the three good solutions
at origin. (We call this possibility a ``conspiracy''.)
One must also be concerned that the regular solution for one spinor
component, say $\beta_0$, might correspond to a solution of other spinor
components which are not well-behaved. However, it is easy to show that
if $\beta_0 \rightarrow 0$ as $r \rightarrow \infty$ then $\gamma_0$, $\beta_{-1}$,
$\gamma_{-1}$ $\rightarrow 0$. Thus, barring conspiracies, we expect to find
only one regular solution of the $(\beta_0, \gamma_0, \beta_{-1}, \gamma_{-1})$
equations.

Numerically solving system (\ref{m=0}), we find a well 
behaved solution (Figure 1). We took $f(r)=\tanh(r)$ and $v(r)= \tanh^2(r)$
for the string profile functions which gives the correct asymptotical behavior
$f \sim r$, $v \sim r^2$ as $r \rightarrow 0$, $f \sim 1$,  
$v \sim 1$ as $r \rightarrow \infty$ and the correct parity. We also set 
$iM_R/(h \eta)=1.1$. Although no other well behaved numerical 
solutions were found, a ``conspiracy" can not be entirely dismissed
because of possible numerical errors. However, more than one neutrino
zero mode would lead to a mismatch between the number of neutrino and
electron zero modes\footnote{An analysis in the case of electrons, 
similar to the one done for neutrinos, shows that there is only one well
behaved solution without the possibility for a conspiracy.}. 
\begin{figure}[!ht] 
\epsfxsize = 0.75 \hsize \epsfbox{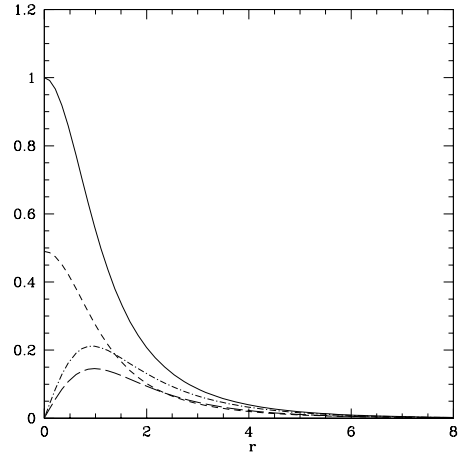} 
\caption{Numerical solution of the system in eq. (\ref{m=0}) showing
$\beta_0$ (solid), $\gamma_0$ (short dash), $\beta_{-1}$ (dot dash)
and $\gamma_{-1}$ (long dash). The solution is for
$iM_R/(h \eta)=1.1$, $f(r)=\tanh(r)$ and $v(r)= \tanh^2(r)$}
\end{figure}

Next we consider non-zero values of the parameter $m$ in 
eq. (\ref{rec}). For the case $m=1$, analysis of large $r$ 
gives results similar to the $m=0$ case --- two exponentially
growing and two exponentially decaying solutions, the latter ones with
the expected large $M_R$ ``see-saw" behavior and the standard 
$M_R \rightarrow 0$ limit. For small $r$, the general well behaved solution is
of the form:
\be \label{m=1sol}
\left( \begin{array}{c} \beta_1 \\ \gamma_1 \\ \beta_{-2} \\ \gamma_{-2} \end{array} \right) 
 = \left( \begin{array}{c}
a_0 r^3 + a_2 r^5 + a_4 r^7 + \ldots \\
b_0 r^1 + b_2 r^3 + b_4 r^5 + \ldots \\
c_0 r^2 + c_2 r^4 + c_4 r^6 + \ldots \\
d_0 r^2 + d_2 r^4 + d_4 r^6 + \ldots 
\end{array} \right) 
\ee
where only two coefficients are independent (say $b_0$ and $c_0$, 
while $a_0=- f_0 h \eta b_0/4$, $d_0= i M_R b_0/4$,
$c_2=- (c_0 v_0 + h\eta f_0 iM_R  b_0 /4 )/2$).
With only two good asymptotic solutions, both at infinity and at the origin, 
we cannot say anything definite about existence of the zero modes in this 
sector. Depending on the specific matching coefficients between small and
large $r$ solutions we could have none, one or two zero modes. However,
numerical analysis shows that there are no nontrivial well-behaved
solutions, i.e. $\beta_1 = \gamma_1 = \beta_{-2} = \gamma_{-2} = 0$ everywhere.
Numerical experiments also indicates that a similar situation occurs 
for $m \ge 2$.

As a simple consequence of the fact that zero modes in $3+1$ dimension 
satisfy the relationship $\omega=\pm k$, they must be eigenstates 
of the $\gamma^0 \gamma^z$ operator.
We found one neutrino zero mode with $\beta \neq 0$, $\gamma \neq 0$ and 
therefore, analysis of the ($\alpha$,$\delta$) pair of the equations of the 
system (\ref{rec}) is not necessary. In order to be eigenstate
of $\gamma^0 \gamma^z$ with the definite eigenvalue
(in this case $-1$) the neutrino zero mode must have $\alpha=\delta=0$.

\section{Conclusion}

We have found that in the standard model 
a massless neutrino has one well-behaved zero mode on a Z-string which,
being part of a continuum, is Dirac delta function normalizable.

We considered the minimal extension of the standard model Lagrangian which
includes a phenomenologically valid massive neutrino, with both Dirac and 
Majorana mass terms. The homogeneous, non-winding Majorana mass term $M_R$ 
makes this case considerably different from the electron one. 
To accommodate the angular dependence of the equations of motion 
we must use an ansatz in which each spinor component is a superposition 
of at least two modes with definite winding number. 
As a consequence, we get a set of coupled equations. 
We solved the equations analytically in two asymptotic regimes, small and large 
$r$, and argued that in each sector determined by a definite
winding number $m$ it is possible, in principle, to have 
one or more zero modes. Numerical analysis indicates that there is
one well-behaved zero mode in the $m=0$ sector and none in
higher $m$ sectors.

\section{Acknowledgements}

One of the authors, Dejan Stojkovic, would like to thank
Zlatko Dimcovic for the help in numerical aspects of this problem.
This work was supported by DOE 3422837 to the Particle/Astrophysics group at
CWRU and NSF 3423593 to GDS.

\end{document}